\documentclass[conference, 10pt]{IEEEtran}
\usepackage{booktabs}
\usepackage[subtle,tracking=normal,bibnotes,charwidths]{savetrees}
\usepackage{amsmath,amssymb,amsfonts}
\usepackage{algorithmic}
\usepackage{graphicx}
\usepackage{textcomp}
\usepackage{xcolor}
\usepackage{hyperref}
\usepackage{siunitx}
\usepackage{comment}

\newif{\ifnotcutspace}
\notcutspacefalse

\newcommand{\dscope}{\textsc{DScope}}
\def\BibTeX{{\rm B\kern-.05em{\sc i\kern-.025em b}\kern-.08em
    T\kern-.1667em\lower.7ex\hbox{E}\kern-.125emX}}
\title{Characterizing the Modification Space \\of Signature IDS Rules}

\makeatletter
\newcommand{\linebreakand}{%
  \end{@IEEEauthorhalign}
  \hfill\mbox{}\par
  \mbox{}\hfill\begin{@IEEEauthorhalign}
}
\makeatother
\author{
    \IEEEauthorblockN{Ryan Guide\IEEEauthorrefmark{1}, Eric Pauley\IEEEauthorrefmark{2}, Yohan Beugin\IEEEauthorrefmark{2}, Ryan Sheatsley\IEEEauthorrefmark{2}, Patrick McDaniel\IEEEauthorrefmark{2}}
    \IEEEauthorblockA{\IEEEauthorrefmark{1}The Pennsylvania State University, \IEEEauthorrefmark{2}University of Wisconsin-Madison
    \\\ \IEEEauthorrefmark{1}rvg106@psu.edu, \IEEEauthorrefmark{2}\{epauley, ybeugin, sheatsley, mcdaniel\}@cs.wisc.edu}
}

\begin{document}
    \maketitle
    \begin{abstract}Signature-based Intrusion Detection Systems (SIDSs) are traditionally used to detect malicious activity in networks. A notable example of such a system is Snort, which compares network traffic against a series of rules that match known exploits.
Current SIDS rules are designed to minimize the amount of legitimate traffic flagged incorrectly, reducing the burden on network administrators. However, different use cases than the traditional one--such as researchers studying trends or analyzing modified versions of known exploits--may require SIDSs to be less constrained in their operation.
In this paper, we demonstrate that applying modifications to real-world SIDS rules allow for relaxing some constraints and characterizing the performance space of modified rules.
We develop an iterative approach for exploring the space of modifications to SIDS rules. By taking the modifications that expand the ROC curve of performance and altering them further, we show how to modify rules in a directed manner. 
Using traffic collected and identified as benign or malicious from a cloud telescope, we find that the removal of a single component from SIDS rules has the largest impact on the performance space.
Effectively modifying SIDS rules to reduce constraints can enable a broader range of detection for various objectives, from increased security to research purposes.\end{abstract}

\pagenumbering{gobble}
\pagestyle{plain}
\section{Introduction}\label{introduction}

Network intrusion detection systems (NIDS) are a component of security that exist to protect networks against adversaries. Signature-based intrusion detection systems (SIDS) are a distinct category of NIDS that utilize known signatures to defend against exploits. While other NIDSs react to anomalous activity, SIDSs have precisely defined rules to identify specific attacks~\cite{liao_intrusion_2013}. An example of one such system is Snort: a free, open source, and lightweight SIDS introduced by Roesch~\cite{roesch_snort_1999}. Snort rules can be written by anyone and the official Snort website~\cite{noauthor_snort_nodate-1} publishes the latest detection rules on a regular basis. As the space of threats continues to evolve, Snort rules are written in response to protect against adversaries. 

Current systems are fine-tuned to minimize false positives, which occur when benign traffic is flagged as malicious. 
Falsely flagging benign traffic is harmful to critical networks, as it disrupts availability and can potentially cause customers to lose trust in the service~\cite{ho_statistical_2012,jajodia_investigating_2008}. As a result, rules have been tailored to fit the semantics of specific attacks, constraining at the same time SIDSs to their traditional security use case. However, if one wanted to achieve different detection objectives, they would not readily be able to do so. The applications of expanding the performance space of SDISs extend to researchers who study specific regions of traffic: whether that be through eliminating all false positives to look at only malicious traffic, or by disregarding false positives and seeking to capture all malicious traffic.

In this paper, we aim to loosen the constraints on SIDS rules through evaluating modifications to the rules. SIDS rules are composed of conditions that must be all met in order for traffic to be matched. Since additions of conditions only serve to further constrain rules, we focus our attention solely on removals. 

Since the full space of removals is intractable, we develop an iterative technique to maneuver within the space for the corresponding desired outcome. We gradually remove parts of the initial rules and compare the receiver operator characteristic (ROC) curves of the modified rules to a heuristic ground truth. We then identify the Pareto frontier~\cite{goodarzi_introduction_2014} of rule performance and repeats our process further until either no further modifications can be made, or the Pareto frontier stops expanding.

We evaluate the modifications made to the rules on the detection performance by using traffic collected and identified as benign or malicious from the \dscope{} cloud telescope~\cite{dscope}. We find the Pareto frontier reachable through removals and identify which and why modifications are the most efficient. We find that single option removal from each rule has the greatest impact, expanding the area under the ROC curve by 11.3\%. Of the removals, we identify the cause of change in detection. 

Through removals, we demonstrate how to loosen constraints on SIDS rules. This enables a broader range of detection for various objectives, from increased security to research purposes.

\section{Background}\label{Background}
An intrusion detection system (IDS) is a system that operates on a network with the intent of blocking malicious traffic while allowing benign traffic to pass through unhindered.
IDSs can be divided into three groups: signature based intrusion detection systems, anomaly based systems, and stateful protocol analysis~\cite{liao_intrusion_2013}. We focus on signature based intrusion detection systems (SIDS).

\paragraph{Snort}\label{Defining Snort}

Snort is an open source SIDS that consists of a collection of rules that are checked against network traffic for matching packets. When a match is found, an alert is triggered~\cite{roesch_snort_1999}. The discovery of new exploits require new detection rules to be written. Our study focuses on Snort 3~\cite{noauthor_snort_nodate-1}. Snort efficiently performs signature and packet comparison in four steps: (1) packet decoding and pre-processing, (2) fast pattern matching, (3) finer rule matching and packet inspection for traffic flagged by previous step, and (4) final logging and triggering of actions (blocking, alerts, etc.). In our study, none of the core functionalities of Snort is modified, only the rules themselves.

\ifnotcutspace
\begin{table}[t]
\footnotesize
    \begin{center}
    \begin{tabular}{ll}
        \toprule
        Option & Usage \\
        \midrule
        \texttt{content} & Matches binary or ASCII literals in traffic\\
        \texttt{sid} & Unique rule identifier \\
        \texttt{flow} & Session properties, e.g., traffic direction\\
        \texttt{detection\_filter} & Only alerts on repeated detection\\
        \texttt{ip\_proto} & Checks protocol number on IP address\\
        \texttt{http\_header} & Limits \texttt{content} to HTTP header\\
        \texttt{dce\_iface} & Checks DCERPC interface \\
        \texttt{isdataat} & Checks for data at a specified location\\
        \texttt{rev} & Which version of the rule is being used\\
        \texttt{byte\_extract} & Creates an integer variable from data\\
        \bottomrule
    \end{tabular}
    \end{center}
    \caption{Rule options in Snort 3 relevant to our study}
    \label{tab:options}
\end{table}
\fi

\paragraph{Snort Rules}\label{Snort Rules}

A Snort rule consists of a header and a body. The header indicates the protocol, source, direction, and destination of the packets to compare against the rule as well as the action to take in case of a match. The body is composed of the different rule options for matching. As soon as a packet fails to meet an option, Snort stops reading the rule ~\cite{roesch_snort_nodate}. \ifnotcutspace We provide an overview of some of the options available in Snort 3 in \autoref{tab:options}.\fi The \texttt{content} option is one of the primary tools used to match packets; the packet is scanned for the pattern specified in the content option. There is no limit as to how many content options can be included in a rule. Other modifiers can also be applied to specify how sensitive the matching should be, and the target section of the packet like \texttt{http\_header}.  These modifiers exist to speed up matching and reduce false positives.

\section{Methodology}\label{Methodology}

The current rules are tailored to specific network attacks to reduce false positives as they can harm availability of critical networks~\cite{ho_statistical_2012,jajodia_investigating_2008}. We seek to make these rules less rigid for other use cases that for instance extend to researchers who study specific regions of traffic: whether that be through eliminating all false positives to look at only malicious traffic, or by disregarding false positives and seeking to capture all malicious traffic.

We are interested in measuring the impact of modifying the rules on the performance of the detection of malicious traffic. We observe that the addition of options to existing rules only further constraint them and decrease detection, as a result, we propose to evaluate the removal of such options. We expect that this will increase the amount of traffic that is detected. 
Among the different rule options, the \textit{general} options--as defined by the Snort documentation~\cite{roesch_snort_nodate}-- have no impact on the performance other than semantics, making removal pointless. In our study, we use a finite set of rules containing 60 distinct options that can be removed without error. We also assume we have access to a data set, discussed in \autoref{Experiments} of benign and malicious traffic to assess the performance of rules. 

\paragraph{The Removal Function}\label{Removal Function}
 
We formally define the body of a rule $R$ as a conjunction of $n$ removable options $a_i$.
\[ R = a_1 \land a_2 \land ... \land a_n = \bigcap_{i=1}^n a_i\]

To gain an understanding of how we are removing options from rules, we define a function $f$ to gather the single removals for each rule: $f(R) = \{R_i ~\mid ~ R_i = R \setminus \{a_i\},~ 0 < i < n\}$
The set of removals can be defined as all $R_i$, where $i$ is index of the option being removed. Since removing all options renders the rule trivial, we omit this possibility. As such, if $R = {a_1}$, $f(R) = \emptyset$.
Generalizing this for more than one removal, we get the function $f(R) = \{R_{i_k} | R_{i_k} = R - \{a_{i_1}, ..., a_{i_k}\}, 0 < i_1 < ... < i_k < n\}$
With $n$ is the number of total removable options, $i$ is the option being removed, and $0 < k < n$ is the number of options being removed. 

\paragraph{Calculating the Space of Modifications}\label{Modspace calculation}
The total number of possible removals can be calculated by counting combinations. Let $i$ be the number of options removed, and let $n$ be the total number of options in a rule. Let $k$ represent all rules, and let $k_i$ be the total number of rules with $i$ removable options, for $0 < i < n$. We then have $k$ represented as 
$k = k_1 \cup k_2 \cup ... \cup k_n$.
We calculate all possible combinations of removals from each subset. For each subset $k_i$, we want to calculate removing $0 < j < i$ options from the rules. 
Thus, we get the following: $k_i = \sum_{j=1}^{i-1} {i \choose j} $
We omit $j = 0$ (the original rules) and $j = i$ (removing all parts).
We calculate each $k_i$ with as $2^i-2$. 
For all $0 < i < n$, we get the following summation  $\sum_{i=1}^{n-1}2^i-2*k_i$

We find that we had over 137 billion possible combinations of removals, which is intractable.
To handle this, we applied some limits to our study, as addressed in the next paragraph. This also enabled us to study the results of removals more closely. We can observe monotonic changes between additional removals. Removing $n$ options at once without knowing the behavior of removing each option individually leads to wild speculation instead of concrete analysis.

\paragraph{Iterative Exploration}\label{Heuristic}

We develop a heuristic to explore iteratively the space of possible removals; to start, we run the original rules on our traffic data and plot them on a ROC curve. We then consider each individual option removal on all the rules, run the traffic detection again on these, and add the corresponding performance points to the ROC curve obtained so far. We then compute the convex hull of all points using the Graham Scan algorithm~\cite{graham_efficient_1972} to obtain the Pareto frontier reachable at that stage. If the Pareto frontier does not expand past the performance of the previous stage, we stop modifications. Otherwise, for each point on the Pareto frontier, we create additionally perform removals on the corresponding rules, and repeat the process described previously. Adopting this heuristic effectively reduces the exploration of modifications to a tractable space, indeed after analyzing the set of $i$ removals from the rules, we no longer have to consider another $n \choose {i+1}$ removal combinations. Instead, by focusing on the $k$ points along the frontier, we perform only $= k {{n - i} \choose 1}$ additional removals.

\section{Evaluation}\label{Experiments}
The Snort rules we use in our experimentation are taken from March 4, 2022 from the Snort website ~\cite{noauthor_snort_nodate-1}. 
To evaluate the performance of rules, we need two key components: traffic to run the rules on and a means to determine which traffic is malicious.

\paragraph{The Rule Set}\label{Rule Study}
Overall we have 42893 rules in our study. There are 60 distinct removable options present in our rules. We show the most commonly occurring options found in our rules in \autoref{tab:rulestats}. The maximum number of options present in any rule is 36, and the minimum is 1. The average number of removable options per rule is 5.93, and the median number of removable options per rule is 6. In over half the rules, the \texttt{content} option is used more than once. We refer to this subset of rules as the \textit{multi-content rules}. 

\begin{table}[]
    \begin{center}
      \begin{tabular}{cc}
      \begin{tabular}{cr}
        \toprule
        Option & Instances \\
        \midrule
        \texttt{content} & 101785 \\
        \texttt{flow} & 42205 \\
        \texttt{service} & 37984 \\
        \texttt{file\_data} & 20833 \\
        \bottomrule
    \end{tabular}
        &
        \begin{tabular}{cr}
        \toprule
        Option & Instances \\
        \midrule
        \texttt{pcre} & 10823 \\
        \texttt{flowbits} & 10330 \\
        \texttt{http\_uri} & 10248 \\
        \texttt{byte\_test} & 4368 \\
        \bottomrule
    \end{tabular}
    \end{tabular}
    \end{center}
    \caption{Number of instances for the most common options.}
    \label{tab:rulestats}
\end{table}

\paragraph{Defining a Ground Truth}\label{Ground Truth}

We need a set of traffic to establish a ground truth, which was done with two data-sets: a collection of traffic from the cloud, and a blocklist of IPs.
Our traffic data comes from the \dscope{} network telescope, which uses public cloud IP addresses to deploy an interactive Internet telescope~\cite{dscope}. 
Network telescopes are ranges of IP addresses used to collect traffic ~\cite{richter_scanning_2019, moore_network_nodate}. In our case, the telescope was set on servers in the AWS US-east-1 region. 
The second data-set determined which traffic was malicious. This set was the fireHOL level 4 blocklist of IPs ~\cite{tsaousis_firehol_level4_nodate}. These IPs were flagged due to scanning activity. 
Our data from the cloud gives us every single IP that communicated with the network over a twenty-four hour period. 
Using the IPs from the cloud, we determine which IPs appeared in the blocklist. Since IPs on the blocklist are considered malicious due to scanning activity rather than using an actual exploit, we consider these IPs benign. This may seem counter-intuitive, but scanning traffic is not inherently malicious. While scans can look for devices to infect, if there is no exploit, then it is not malicious. Hence Snort would not flag it as malicious. The IPs monitored were not allocated, so it follows that the traffic was not for legitimate services and can be classified as malicious ~\cite{dscope}. We gather the source IP addresses from collected traffic. This yields \SI{43.3}{k} IP addresses communicating with our telescope. Using the blocklist, we find \SI{37.6}{k} malicious IPs and \SI{5.6}{k} benign ones.

\paragraph{Creating a Pipeline}\label{Pipeline}

In order to efficiently analyze large sets of data, we create a pipeline to manage the flow of data and speed up the entire process. The pipeline consists of four phases: parsing, pre-processing, rule selection, and output.
The pipeline runs in the following phases:
\begin{itemize}
    \item Parsing: Using original rules, our parser generates every possible removal of $n$ options, then filters down to the desired subsets for examination. The rules are then run on the cloud data.
    \item Pre-processing: The output from Snort is parsed to create an efficient mapping of the traffic matched by the new rules.
    \item Rule Selection: Given an option, the selector uses mappings to identify rules that have that option removed along with the original rules. This enables the analysis of specific option removals. Not giving an option results in all rules being used.
    \item Output: The pipe identifies which source IPs were found by the rules selected in the previous phase alongside the original rules. The blocklist is used to determine the rates of true and false positives.
\end{itemize}

\paragraph{The Original Rules}\label{Original Rule Analysis}

Using the ground truth from the FireHOL blocklist, we identify where the original set of rules are on the ROC curve. For the original set of rules, we find that the true positive rate is 44.54\%, and the false positive rate is 7.79\%. 

\paragraph{The Pareto Frontier}\label{Q1} 
We run four iterations of removals. For each option that is removed, we run the modified rules with that option removed alongside the rules unaffected by that removal. For example, to test the removal of \texttt{http\_header}, we run all the original rules that did not have \texttt{http\_header} alongside the modified rules that have \texttt{http\_header} removed. Through that configuration we observe the impact of each removal. We study the overall performance of each modification rather than each individual rule. This enables a better understanding of the rules in general, whereas working with rules at the individual level would only allow us to study specific cases. Specific cases are uninteresting in this study as they are too narrow in scope. Running the entire subset together achieves the same result as summing the results of each individual rule.

\begin{figure}[!ht]
    \centering
    \includegraphics[width=0.8\columnwidth]{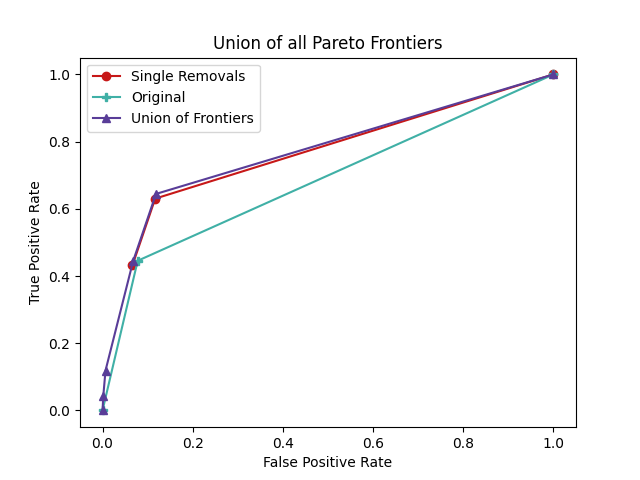}
    \caption{ROC curves for the original rules, single removals, and the union of all Pareto frontiers.}
    \label{fig:Final Frontiers}
\end{figure}

After plotting each removal, we calculate the Pareto frontier for all the results together to determine if the frontier expanded. At each iteration we calculate the area under the Pareto frontier, as shown in \autoref{tab:Area Scores}. We see that the greatest expansion comes from the removal of a single option, and once it reaches the fourth iteration, the expansion stops. The first iteration expands the frontier by 11.3\% of the original size. The second iteration increases the Pareto frontier by only 1.14\%. Due to the small change in difference, we primarily focus our analysis on the first iteration. The original frontier, single removals, and the final curve are all plotted in \autoref{fig:Final Frontiers}

The points along the Pareto frontier from the fourth iteration were all less than or equal to the frontier points from the third iteration. With the exception of removing \texttt{http\_header}, \texttt{http\_uri}, and \texttt{flowbits}, the points with lower true and false positives than the original rules were all due to inverting bad classifiers. We discuss the cause for the change in detection in detail later on. \textbf{Summary:}
\begin{itemize}
    \item The greatest change in the frontiers comes from the removal of a single option.
    \item The Pareto frontier stops expanding after three removals.
\end{itemize}

\begin{table}
\begin{center}
\begin{tabular}{cc}
     \toprule
    Removals & Area of Pareto Frontier(\%) \\
    \midrule
    Original (0) & 68.37\\
    Single (1) & 76.10\\
    Double (2)  & 76.97 \\
    Triple (3)  & 76.99 \\
    Quadruple (4) & 76.99 \\
    \bottomrule
\end{tabular}
\end{center}
\caption{\label{tab:Area Scores}The area covered by each ROC curve as a percentage}
\end{table}

\paragraph{The Options that Caused Movement}\label{Q2} 
Breaking down our Pareto frontiers further, we look at the specific removals that caused the frontiers to change.
To understand the meaning behind the performances, it is necessary to look at the costs of both false positives and false negatives. A false positive disrupts legitimate traffic, while a false negative allows malicious traffic into the network. To measure the trade off between the two, we define a utility function. Let $\theta \in [0,1]$ be the trade off between false positives ($f_p$) and false negatives ($f_n$). We define a cost function, $C$, as $C = \theta f_p + (1-\theta)f_n$
When $\theta \in [0,0.5)$, reducing false negatives is prioritized. Conversely, when $\theta \in (0.5, 1]$ reducing false positives is prioritized. Naturally $\theta = 0.5$ is equivalent to assigning equal weight to both.

\begin{figure}[!ht]
    \centering
    \includegraphics[width=0.49\textwidth]{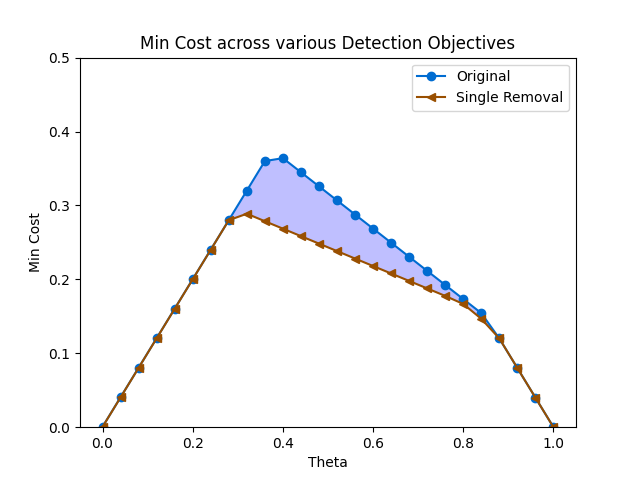}
    \caption{Minimum cost to achieve ideal detection for various costs on false positives and false negatives for the original rules and single removals. The shaded area represents the area reduced by the modifications.}
    \label{fig:Min Cost}
\end{figure}

Looking at \autoref{fig:Min Cost}, we see that the minimum cost for balancing false positives and negatives was lower for our set of single removals than the original rules. Up until $\theta = 0.3$, the costs are equivalent. At that point the minimum cost begins to decrease for single removals, but continues to increase for the original rules. At $\theta = 0.87$, the two lines converge again. The area under the curve for the original rules is 0.2064, and the area under the curve for single removals is 0.1792, giving us a total decrease in area by 0.0272. This illustrates how our modifications reduce the overall cost of false positives and negatives.

At both $\theta = 0$ and $\theta = 1$ we see a cost of 0. When $\theta = 0$, the recommended course of action is to block all traffic. Conversely, at $\theta = 1$, the optimal move is to allow all traffic. These are extreme cases and are not likely to be used by enterprises. 

For $\theta \in (0,0.74)$, the modification required to achieve the minimum cost for single removals is to remove a \texttt{content} option from multi-content rules. For $\theta \in [.74,.85)$, the optimal modification is to remove \texttt{http\_header}. For all other values of $\theta$, the ideal modification lays on the line between the origin and the removal of \texttt{http\_header}. 

For the original rules, we find that the minimum cost is achieved at the original rules exactly for $\theta \in [0.38, 0.84]$. We note that this correlates with the values of $\theta$ where we were able to reduce the minimum cost.

For any given detection objective, our rules perform at least as well as the original rules, with the greatest difference occurring from $\theta \in [0.30,0.70]$. This means that our rules had a lower minimum cost than the original rules when relatively equal weights were applied to the costs of false positives and negatives.

To analyze the impact of each modification, we examine the precision, recall, and the f1-score of single removals ~\cite{powers_evaluation_nodate}. Recall is the percentage of malicious IPs that were detected by a rule set out of the entire malicious IP space. Precision refers to the percentage of malicious IPs found in the set of IPs identified by our rules. The f1-score is a metric that takes the harmonic mean of precision and recall ~\cite{sasaki_truth_nodate}. The f1-score is used as a combination of both precision and recall. Given our class imbalance, we consider the precision, recall, and f1-score for both malicious and benign traffic classification. The top f1-scores are shown in \autoref{tab:F1-Scores}. 

To understand why we look at both metrics, we consider the removal of the \texttt{flags} option. Upon removal, we find that every single IP address in the space is matched. We call this behavior \textit{universal matching}. Calculating precision and recall for malicious traffic detected gives us a precision of 86.96\% and a recall of 100\%. However, when we examine benign traffic precision and recall, we get 0\% for both. The f1-scores for malicious and benign traffic are 0.9302 and 0, respectively. This demonstrates that while removing \texttt{flags} looks like an excellent choice for detecting malicious traffic, we find that it is useless in terms of correctly classifying traffic as benign. By looking at both scores we can determine which removals are actually worthwhile. We also use these scores to find the macro f1-score ~\cite{opitz_macro_2021}. 

First, we look at the original rules. We examine the malicious traffic metric first. The recall is only 44.54\%, but the precision is 97.44\%, indicating that while it misses roughly half the malicious traffic, it is quite precise. This provides weight to our assertion that the rules are geared towards preventing false positives. We get an f1-score of 0.6113. The cause for this score comes from the lower true positive rate. Looking at the metric for benign traffic, we find a precision of 19.96\% and a recall of 92.20\% resulting in an f1-score of 0.3282. This is lowered significantly by the low precision. We find a macro f1-score of 0.4697. We find that many of the modifications have very similar scores.
We direct our focus in this section to removals from our Pareto frontier and removals with a high macro f1-score.

We now examine notable points from single removals. Here we consider the highest macro f1-scores as well as the points on the frontier for single removals. For each option, we examine the score and factors that influence it. A high f1-score can be misleading, given that most removals saw a high precision for malicious traffic.

Looking at the removal of \texttt{http\_header}, we see that the recall drops by 1.3\%, and the precision increases by .3\%. As a result, the f1-score drops to 0.5988. The minor improvement in the ratio of true and false positives comes at a greater cost to recall, and thus worsens the f1-score. We also see a decrease in the f1-score for benign traffic, resulting in an overall drop in the macro f1-score.

Removing a single \texttt{content} option from multi-content rules had 63.03\% true positive rate and a false positive rate of 11.67\%. The precision of this set decreased by .15\% from the original rules. The f1-score was 0.7650 for this set, which reflects the higher recall than the original rules. 
The removal of \texttt{content} from rules regardless of remaining \texttt{content} options did increase the recall by over 50\% to 99.78\%, but at a cost to precision of over 10\%, bringing it down to 86.93\%. This resulted in a f1-score of 0.9292. The increase in f1-score comes entirely from the surge in recall. It is important to note that removing \texttt{content} resulted in matching 100\% of false positives, and since the false positives are greater than the true positives, we would invert this point to achieve a good classifier. This inverted classifier would have a precision of 100\%, but a small recall of 0.22\%.

The removal of \texttt{detection\_filter} had the second highest macro f1-score of 0.5567. Removing \texttt{detection\_filter} gave us a true positive rate of 63.83\% and a false positive rate of 26.91\%. The increase in true and false positives was 19\% for both, indicating a linear increase rather than a bias towards one or the other, hence this point alone did not change the Pareto frontier alone. When combined with the removals of \texttt{flow} and \texttt{dce\_iface}, it created a bad classifier with a higher false positive rate than true positive rate. Upon inverting this classifier, we obtained a new frontier point.

Removing \texttt{ip\_proto} had the third highest macro f1-score, and was the last single removal to achieve a macro f1-score over 0.5. The score was 0.5169. The true and false positive scores for the removal of \texttt{ip\_proto} were 52.81\% and 11.89\% respectively. This was an increase in true positives by 8\%, and an increase in false positives by 4\%. This improvement is eclipsed by the removal of \texttt{content}, and therefore is not included in the frontier for single removals. Similarly to \texttt{detection\_filter}, it does appear on the union of all frontiers. 

The combination of removing both \texttt{dce\_iface} and \texttt{flow} resulted in a bad classifier, but the inverse expanded the Pareto frontier. The inverted removal of both resulted true and false positive rates of 11.85\% and 0.66\% respectively. The individual removals of \texttt{flow} and \texttt{dce\_iface} were not part of the frontier for single removals. We do note that the removal of \texttt{dce\_iface} was the cause for the bad classification, as \texttt{flow} found more malicious traffic than benign. The f1-scores for both malicious and benign traffic were both quite low, at 0.2117 and 0.2525 respectively. The macro f1-score was 0.2321, which is half the score of the original rules. When \texttt{detection\_filter} was included the inverted classifier fell to true and false positive rates of 4.19\% and 0.16\%. The macro f1-score was 0.1593, lowered entirely by the drop in true positives. \textbf{Summary:}
\begin{itemize}
    \item We showed using our utility function that the minimum cost for our modifications is at least as good as the original rules, with the greatest decrease in cost coming from assigning roughly equal costs to false positives and negatives.
    \item Removing a single \texttt{content} from multi-content rules was the most effective modification in terms of macro f1-score.
    \item While \texttt{detection\_filter} and \texttt{ip\_proto} increased detection, we show that there are better alternatives.
\end{itemize}

\begin{table}
\begin{center}
\begin{tabular}{cccc}
    \toprule
    Removal & $F1_{\text{malicious}}$ & $F1_{\text{benign}}$ & $F1_{\text{macro}}$ \\
    \midrule
    \texttt{content} (when $>1$) & 0.7650 & 0.4063 & 0.5857 \\
    \texttt{detection\_filter} & 0.7605 & 0.3549 & 0.5567 \\
    \texttt{ip\_proto} & 0.6832 & 0.3506 & 0.5169 \\
    \texttt{flags} & 0.9302 & 0.0 & 0.4651 \\
    original rules & 0.6113 & 0.3282 & 0.4697 \\
    \bottomrule
\end{tabular}
\end{center}
\caption{\label{tab:F1-Scores}F1-Scores for significant removals.}
\end{table}

\paragraph{The Cause of Movement}\label{Q3} 
In this section we explain why certain modifications alter detection in the manner that they do and the value in removing each option.
The \texttt{flags} option checks TCP flags on the packet. Without this filter traffic more traffic is matched. This increase is arbitrary, and is not particularly insightful. There are only 24 rules in which \texttt{flags} was present: less than 1\% of all our rules.

The decrease in detection from \texttt{http\_header} can be explained by the Snort 3 documentation. Since the header is not decoded the detection cursor might not find the matching string. There are 3434 rules with \texttt{http\_header} present in our rules, meaning the removals affect only 8\% of the rules. We see a slight decrease in detection from the removal of \texttt{http\_header} where the decrease in true positives is slightly larger than the decrease in false positives.

Next we consider the \texttt{content} option. As demonstrated, a subset of \texttt{content} proved to be one of the optimal removals. The change in \texttt{content} can be split into two categories; removing \texttt{content} in general and the subset of multi-content rules. Since \texttt{content} is the primary tool used for detection, removing it enables more matches. Removing \texttt{content} from a multi-content rules greatly increased the range of detection. Since we saw nearly complete matching of all IPs, we classify this increase as random. Even though the inverse of this results in obtaining a small percentage of true positives with no false positives, it is likely this is random noise, and therefore is not taken into consideration.
Examining the subset of \texttt{content} removals from multi-content rules yields a significantly lower rate of false positives. We instead see that the false positive rate is 11\%, while the true positives increased to 63\%. As there are still \texttt{content} options left in the rule, traffic that matches less of the characteristics is collected. However, this enables capturing of similar malicious traffic. A slight adjustment to the malicious traffic, either repositioning where the exploit is in the payload of the packet or an alteration to the exploit itself can be missed by the original rule. When the \texttt{content} option to match that part of the exploit is no longer present, the rule becomes more flexible in detecting variations on the attack.
The other key benefit to having more than one \texttt{content} option is the repetition it enables. Consider a rule with three \texttt{content} options: $A$, $B$, and $C$. Removing a single \texttt{content} option gives us three new rules: one with $A$ and $B$, one with $B$ and $C$, and one with $A$ and $C$. Compared to the original rule, we now have three rules which are able to sense variants on the original attack.
Similar behavior was observed in another study \cite{aickelin_rule_2007}, where they replaced characters in the \texttt{content} option string with general characters. From this they found variations on attacks that were missed prior.
Out of the 42893 rules present in our experiment, 24769 of them are multi-content rules. This means about 57\% of the rules are modifiable in this manner. There are only 344 rules without any \texttt{content} options, meaning 99\% of all our rules has at least one \texttt{content} option.

The \texttt{detection\_filter} option adds a counter to the rule, indicating that it must receive a certain number of communications from an IP before it will allow the rule to fire. Thus removing this logically eliminates the filter and allows the rules to fire more often. As a result, we found that half of the additional traffic flagged was malicious, giving us an almost equal increase. There are 279 rules (less than 1\% of the rules) that have a \texttt{detection\_filter} option.

The \texttt{ip\_proto} option checks the protocol number on the IP address. This number is used to determine the protocol on the next level of the network. The increase in detection is twice as large for malicious traffic as it was for benign traffic. While removing the protocol number picked up some random traffic, it is possible the removal identified variants of the malicious traffic that used new protocol numbers. Only 33 rules have the \texttt{ip\_proto} option.

The option \texttt{dce\_iface} checks DCERPC (distributed computing environment remote procedure calls) interface.  
The \texttt{dce\_iface} option is designed to normalize the universally unique identifier that clients use to communicate with servers under this protocol. The Snort 3 documentation describes this option as a means to eliminate false positives from multiple services being connected to the server ~\cite{roesch_snort_nodate}. It then follows that removing this option increases false positives at a greater rate than true positives, giving us a bad classifier for single removals. Only 235 rules have the \texttt{dce\_iface} option.

The combinations that form the last Pareto frontier are heavily influenced by the single options within each group. The jump in false positives seen in \texttt{dce\_iface} and \texttt{flow} comes from the removal of \texttt{dce\_iface}. The removal of \texttt{content}, \texttt{ip\_proto}, and \texttt{isdataat} is likely a random increase, rather than a useful modification. \textbf{Summary:}
\begin{itemize}
    \item The changes caused by removing \texttt{detection\_filter}, \texttt{flags}, and \texttt{http\_header} are random.
    \item Removing \texttt{dce\_iface} results in a much higher rate of false positives.
    \item The most effective modification is removing \texttt{content} from multi-content rules, due to the increased sensitivity to variants to exploits.
    \item The options \texttt{flow} and \texttt{ip\_proto} only appear on the Pareto frontier when paired with other removals.
\end{itemize}

\paragraph{Effective Modifications}\label{Effective Modifications}
Based on the macro f1-scores, removing \texttt{content} from multi-content rules is the most effective modification. With the size of this subset, we find that 57\% of the rules are candidates for effective modifications. For achieving stronger defense at lowest cost to false positives, removing a single \texttt{content} from multi-content rules is the most effective modification. We see this demonstrated not only in the macro f1-scores, but also from the utility function. For reducing false positives at the cost of detection, the ideal modifications are inverted results of removing \texttt{flow} and \texttt{dce\_iface}.

\section{Conclusion}\label{Conclusion}

Characterizing the performance space of SIDS enables novel approaches to studying network traffic. Through modifications, SIDS rules can be used to collect various sets of traffic based on the desired outcome. We demonstrated that through the most effective removal of a single \texttt{content} option from rules with at least two content options, the performance space expands by 11.3\%. We distinguished the difference between effective modifications and modifications that result in random changes detection. Through these modifications, SIDS rules can be used for other purposes than just network intrusion detection systems.

\section*{Acknowledgments}
This material is based upon work supported by, or in part by, the National Science Foundation under Grant No. CNS-1805310, and the U.S. Army Research Laboratory and the U.S. Army Research Office under Grant No. W911NF-13-2-0045. Any opinions, findings, and conclusions or recommendations expressed in this publication are those of the author(s) and do not necessarily reflect the views of the National Science Foundation, or the U.S. Government. The U.S. Government is authorized to reproduce and distribute reprints for government purposes notwithstanding any copyright notation hereon.

\bibliographystyle{IEEEtran}
\bibliography{refs.bib}

\end{document}